\documentclass[aip,apl,reprint,amsmath,amssymb]{revtex4-1}

\usepackage{graphicx}
\usepackage{dcolumn}
\usepackage{bm}

\usepackage[utf8]{inputenc}
\usepackage[T1]{fontenc}
\usepackage{mathptmx}
\usepackage{etoolbox}
\usepackage{xcolor}

\draft 

\makeatletter
\def\@email#1#2{%
 \endgroup
 \patchcmd{\titleblock@produce}
  {\frontmatter@RRAPformat}
  {\frontmatter@RRAPformat{\produce@RRAP{*#1\href{mailto:#2}{#2}}}\frontmatter@RRAPformat}
  {}{}
}%
\makeatother

\begin{document}


\title{Polarization-resolved measurement of forward volume spin waves by micro-focused Brillouin light scattering} 



\author{Krzysztof~Szulc}
\email{krzysztof.szulc@ceitec.vutbr.cz, michal.urbanek@ceitec.vutbr.cz}
\affiliation{ 
CEITEC BUT, Brno University of Technology, Brno 612 00, Czechia
}%
\affiliation{ 
Institute of Molecular Physics, Polish Academy of Sciences, Poznań 60-179, Poland
}%
\author{Mengying~Guo}
\affiliation{ 
School of Physics, Hubei Key Laboratory of Gravitation and Quantum Physics, Institute for Quantum Science and Engineering, Huazhong University of Science and Technology, Wuhan, China
}%
\author{Ond\v{r}ej~Wojewoda}
\affiliation{ 
CEITEC BUT, Brno University of Technology, Brno 612 00, Czechia
}%
\affiliation{ 
Department of Materials Science and Engineering, Massachusetts Institute of Technology, Cambridge, MA, USA
}
\author{Hongyu~Wang}
\affiliation{ 
Beijing National Laboratory for Condensed Matter Physics, Institute of Physics, Chinese Academy of Sciences, Beijing, China
}%
\author{Dominik~Pavelka}
\affiliation{ 
CEITEC BUT, Brno University of Technology, Brno 612 00, Czechia
}%
\author{Jan~Kl\'{i}ma}
\affiliation{ 
CEITEC BUT, Brno University of Technology, Brno 612 00, Czechia
}%
\author{Jakub~Kr\v{c}ma}
\affiliation{ 
CEITEC BUT, Brno University of Technology, Brno 612 00, Czechia
}%
\affiliation{ 
Institute of Physical Engineering, Brno University of Technology, Brno 616 69, Czechia
}%
\author{Xiufeng~Han}
\affiliation{ 
Beijing National Laboratory for Condensed Matter Physics, Institute of Physics, Chinese Academy of Sciences, Beijing, China
}%
\author{Qi~Wang}
\affiliation{ 
School of Physics, Hubei Key Laboratory of Gravitation and Quantum Physics, Institute for Quantum Science and Engineering, Huazhong University of Science and Technology, Wuhan, China
}%
\author{Michal~Urb\'{a}nek}
\affiliation{ 
CEITEC BUT, Brno University of Technology, Brno 612 00, Czechia
}%
\affiliation{ 
Institute of Physical Engineering, Brno University of Technology, Brno 616 69, Czechia
}%

\date{\today}

\begin{abstract}
We show how the micro-focused BLS signal of forward volume spin waves is formed and why it remains observable despite symmetry-based “suppression” expectations. A reciprocity-theorem based model with vectorial diffraction-limited focusing identifies the nonnegligible longitudinal focal-field component as the key element responsible for BLS sensitivity in the forward volume geometry. We further demonstrate that full polarization analysis, implemented through polarizer–-analyzer maps of coherently excited spin waves, provides information beyond the conventional crossed polarizer-–analyzer readout. In a BiYIG thin film, the measured maps exhibit Stokes/anti-Stokes polarization asymmetries and nontrivial patterns that stem from quadratic magneto-optical coupling terms. Fitting the data with a model including Voigt and Cotton–-Mouton contributions yields an effective Cotton-–Mouton constant and shows that the quadratic response is comparable to the linear Voigt contribution.
\end{abstract}

\pacs{}

\maketitle 


Brillouin light scattering (BLS) is among the most versatile optical methods of measuring condensed matter excitations, providing access to their frequency, wavevector, and phase via inelastic scattering of laser light. In magnetic systems, BLS directly measures thermal or externally driven spin waves (magnons) through Stokes (magnon creation) and anti-Stokes (magnon annihilation) processes.\cite{madami2012application,kargar2021advances}
From the experimental point of view, two BLS implementations dominate in magnonics research: (i) conventional, $k$-resolved BLS, where the in-plane magnon wavevector is selected by the scattering geometry; and (ii) micro-focused BLS (\textmu{}BLS), where an objective with a high numerical aperture (NA) focuses the probing laser beam to a diffraction-limited spot, enabling spatially resolved mapping of spin-wave intensity and phase.\cite{sebastian2015micro} There are also recent nanophotonic approaches which further extend \textmu{}BLS capabilities and allow interactions of spin waves with free space inaccessible light.\cite{Jersch2010, Freeman2020, wojewoda_observing_2023, wojewoda_phase-resolved_2023, krcma2025mie}.

For a long time, the \textmu{}BLS community had the intuition that with the optical axis normal to a magnetic film, the detection of spin waves in the forward volume (FV) configuration (uniform out-of-plane magnetization) is symmetry suppressed. In the simplified picture, the longitudinal focal-field component $E_z$ was often assumed negligible for rotationally symmetric illumination (analogous to the common paraxial neglect of longitudinal-field contributions\cite{novotny2006}), and the remaining transverse-field contribution was expected to cancel: in the FV geometry the dynamics is purely in-plane and the associated magneto-optical (MO) response carries an odd azimuthal symmetry that averages to zero when coherently integrated over the full aperture.

The key change in perspective comes from a full electromagnetic description of the focusing: a high-NA objective does not generate a purely transverse field in the focal region. Instead, the focused field contains all three components ($E_x,E_y,E_z$), and the longitudinal component can be only a few times smaller than the dominant transverse component(s), rather than being negligible.\cite{wojewoda2024modeling, benaziz2025method} Consequently, the observability of FV spin waves depends on correctly accounting for the vectorial structure of the focused field and the subsequent MO coupling that transforms the incident field into the scattered field collected by the objective.

Experimentally, spin-wave modes in out-of-plane (or nearly out-of-plane) magnetized films have in fact been observed with \textmu{}BLS in multiple contexts,\cite{merbouche2022giant,wang2023deeply} underscoring that FV detection is practically achievable even if certain idealized symmetry arguments predict its suppression. The modeling viewpoint therefore reframes the question: not “can spin waves in the FV geometry be detected?”, but rather “which MO tensor terms and which electric field components dominate the FV signal under a given optical and magnetic configuration?”

Another key aspect of magnon BLS is the polarization selectivity of the MO scattering process. In the simplest (and most commonly used) description, the polarization of the magnetically scattered light is rotated by $90^\circ$ with respect to the incident polarization\cite{sebastian2015micro}, which motivates the widespread use of crossed polarizer--analyzer detection to suppress the elastically scattered background and enhance the magnon signal. Most experiments record only the intensity projected onto the polarization orthogonal to the incident beam, rather than capturing the full polarization state of the scattered field.

Here we show that the polarization state of the scattered light contains additional information about the underlying MO coupling. This opens a route to use polarization not only as a filter but as an additional information carrier: by measuring full polarizer--analyzer maps and by employing modeling of the \textmu{}BLS signal, one can access and distinguish linear and quadratic contributions to the MO coupling.


\begin{figure}
    \includegraphics[width=\linewidth]{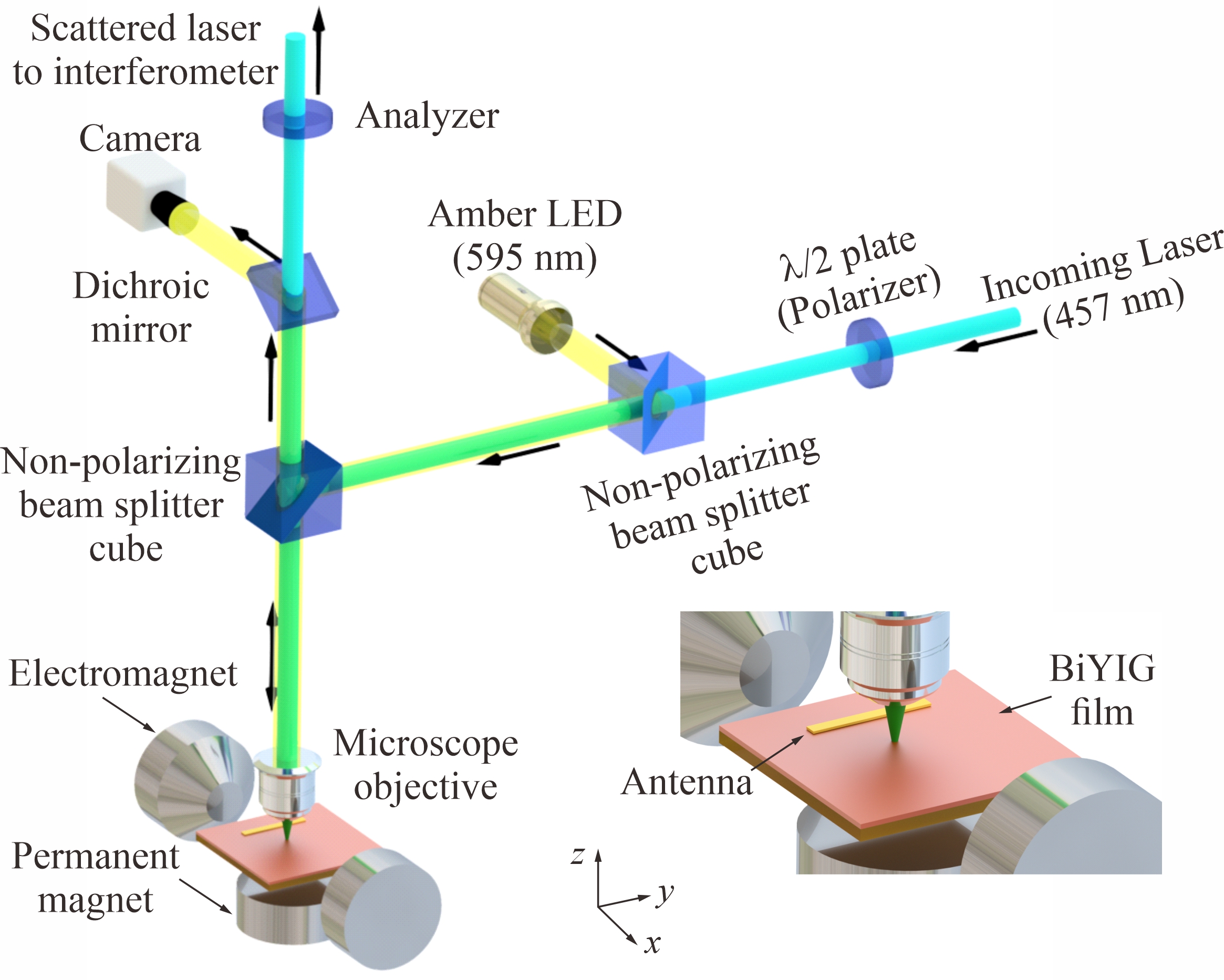}
    \caption{\label{fig:setup} Schematics of the used micro-focused Brillouin light scattering setup. The polarization of the linearly polarized incoming laser can be rotated by motorized $\lambda/2$ plate. The beam then passes through two non-polarizing beam splitters and is focused on the sample by microscope objective with NA\,=\,0.8. The scattered light then passes through a rotatable polarizer before is fed into a tandem Fabry--Perot interferometer. An additional LED light source and CCD camera are used to control the positioning of the laser beam on the sample in the vicinity of the microstrip antenna. The biasing magnetic field can be applied both, in-plane and out-of-plane.}
\end{figure}

We have investigated a 100\,nm-thick bismuth-doped yttrium iron garnet (BiYIG) thin film with the stoichiometric composition Bi\textsubscript{1}Y\textsubscript{2}Fe\textsubscript{5}O\textsubscript{12}. The film was deposited on a (111)-oriented Gd\textsubscript{3}Ga\textsubscript{5}O\textsubscript{12} (GGG) substrate using radio-frequency (RF) magnetron sputtering.\cite{wang2026ultrathin}

The BLS experiments on coherently excited spin waves were performed in the \textmu{}BLS setup shown in Fig.~\ref{fig:setup}. A single-frequency laser with a wavelength of 457~nm is focused on the sample using a microscope objective with NA~=~0.8, giving a laser spot size of about 350~nm. The polarization angle of the incident (linearly polarized) laser beam is controlled by a motorized half-wave plate (hereafter, the polarizer). The beam then passes through two non-polarizing beam splitters and is focused on the sample by the microscope objective. The scattered light then passes another non-polarizing beam splitter and is analyzed by a motorized rotating polarizer (hereafter, the analyzer). After passing the analyzer, the laser beam is then spectrally analyzed in a tandem Fabry--Perot interferometer (TFPI) with a single-photon detector \cite{Lindsay1981}. The bias magnetic field can be applied either in the sample plane using an electromagnet\,---\,for Damon–Eshbach (DE, $\mathbf{k}\perp\mathbf{m}$) and backward volume (BV, $\mathbf{k}\parallel\mathbf{m}$) geometries\,---\,or out of the plane using a permanent magnet for the FV geometry. The coherent spin waves were excited by a 2~\textmu{}m-wide microstrip antenna connected to an RF generator. The DE and BV spin waves were linearly excited by the spatially varying out-of-plane magnetic field in the region outside of the antenna. In case of FV spin waves, a frequency-shifted nonlinear FMR mode was excited by strong in-plane magnetic field under the antenna and the spin waves gained non-zero momentum in the region outside of the antenna as the dispersion returned into the linear regime \cite{wang2023deeply}. 

We started with experiments in the FV geometry. The sample was placed on a permanent magnet providing an out-of-plane field of 330\,mT. We fixed the RF excitation at the frequency of 6.36\,GHz (at maximum measured BLS intensity, corresponding to $k\approx10$ rad/\textmu{}m) and rotated both the polarizer and analyzer. For each polarizer--analyzer configuration ($6^\circ$ step), we acquired a full BLS spectrum, obtaining a complete 2D dataset. Next, we acquired the full 2D polarizer--analyzer datasets for DE and BV geometries. We set the in-plane field to 25\,mT and swept the RF frequency to find the maximum BLS intensity (2.52 GHz, corresponding to $k=0$). Then we shifted the frequencies by $\pm$75\,MHz and measured DE mode at 2.59\,GHz and BV mode at 2.44\,GHz, these frequencies correspond to the wavevector $k\approx0.7$ rad/\textmu{}m).

\begin{figure}[t]
    \includegraphics{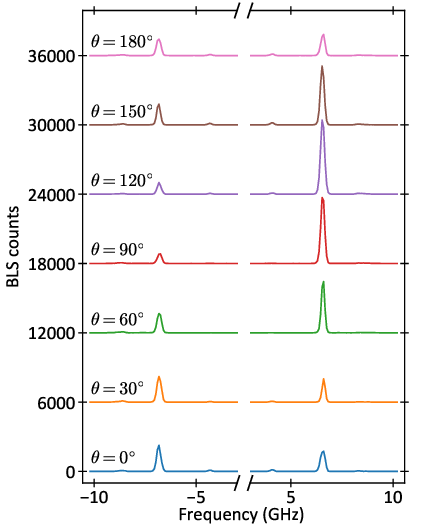}
    \caption{\label{fig:spectra} Measured \textmu{}BLS spectra in the FV geometry for 7 different polarizer angles $\theta$ with the analyzer angle set to $\phi = 150^\circ$. Spectra are offset by 6000 counts from each other for clarity.}
\end{figure}

A selected set of BLS spectra extracted from the full FV dataset is shown in Fig.~\ref{fig:spectra}. The spectra were acquired at 7 different polarizer angles $\theta$ when the analyzer angle was set to $\phi=150^{\circ}$. The primary peaks are close to the frequency of 6.36~GHz excited by the antenna and RF generator. As expected, the amplitudes of the primary peaks significantly change with the polarizer angle. In addition to the main peaks, there are two smaller peaks present symmetrically on both sides of the spectrum. The first peak lies at a frequency of about 4~GHz and can be attributed to a spurious elastically scattered laser mode. This peak is important because it can serve as a polarization reference: since it originates from elastically scattered light, its polarization remains unchanged, and its maximum amplitude is always obtained when the polarizer and analyzer are parallel. We used this peak to calibrate the relative polarizer--analyzer position; the calibration procedure is described in Supplementary Material. The second peak, at about 8~GHz, has the character of a magnon excitation and can be attributed to a higher-order spin-wave mode. It could be analyzed by applying the same procedure as we used for the analysis of the primary peak but this analysis goes beyond the scope of this paper. The following analysis focuses only on the primary peaks observed at the frequencies of 6.36~GHz, 2.59~GHz, and 2.44~GHz for FV, DE, and BV fundamental spin-wave modes, respectively. 

Looking at the individual BLS spectra in Fig.~\ref{fig:spectra}, we can immediately see that the maximum peak heights are measured at different polarizer angles $\theta$ for the Stokes (max. at $\theta=0^\circ$) and anti-Stokes (max. at $\theta=120^\circ$) scattering processes. As we show later, this property indicates a significant contribution of a higher-order MO effect on a scattered laser beam. The quadratic Cotton--Mouton effect is known to be present in garnets such as YIG, where its strength can be comparable with the linear Kerr effect.\cite{cottam1976theory,grunberg1977light,camley1978light,dutcher1989brillouin,cochran2000brillouin,giovannini2001theory}

To interpret BLS spectra across different geometries and to analyze the polarization-dependent datasets, we modeled the experimental setup using the open-source Python package SpinWaveToolkit.\cite{swt, klima2026} This package can calculate spin-wave dynamics using Kalinikos--Slavin model\cite{kalinikos1986theory} and combine it with semi-analytical model of the BLS-signal formation.\cite{wojewoda2024modeling, krcma2025mie} The BLS-signal calculation is carried out in three steps. First, for a given geometry and excitation frequency, we calculate the wavevector-dependent Bloch function using the Kalinikos--Slavin model,\cite{wojewoda2024modeling} which represent the dynamic magnetization of the spin wave. This step requires the spin wave to be in the linear regime. We assume that since the detection in the FV geometry is done far away from the antenna, the spin wave is already in the linear regime, while in the BV and DE, the amplitude of the excitation does not exceed the nonlinearity threshold. Next, the dynamics is influenced by the MO coupling, creating the dynamic susceptibility tensor. Finally, the susceptibility is coupled to the optical problem: it enters the MO scattering formalism and determines the polarization-resolved inelastic light scattering intensity collected by the objective. The full susceptibility tensor $\boldsymbol{\chi}$, which incorporates both the linear Kerr contribution and the quadratic Cotton--Mouton term, can be written as
\begin{equation}\label{eq:chi}
    \chi_{ij} = iQ\,\varepsilon_{ijk}\,m_k + B_{ij}\,m_i m_j
    \qquad (i,j,k\in\{x,y,z\}),
\end{equation}
where $m_{i}=M_{i}/M_{\rm s}$ is the unitless $i$-component of the magnetization vector, $Q$ is the Voigt constant, $B_{ij}$ are the Cotton--Mouton constants, and $\varepsilon_{ijk}$ denotes the Levi-Civita symbol. For the calculation of the dynamic effects, the MO susceptibility tensor needs to be linearized with respect to the static magnetization. For the FV geometry, the resulting linearized tensor reads
\begin{multline}\label{eq:chi_linear}
    \boldsymbol{\chi} = 
    \begin{pmatrix}
        0 & 0 & -iQm_y \\
        0 & 0 & iQm_x \\
        iQm_y & -iQm_x & 0
    \end{pmatrix}
    +\\+
    \begin{pmatrix}
        0 & 0 & B_{xz}m_x \\
        0 & 0 & B_{yz}m_y \\
        B_{xz}m_x & B_{yz}m_y & 0
    \end{pmatrix}.
\end{multline}
An analogous procedure is used for the DE and BV geometries.

In the next step, we calculate the intensity of the BLS signal collected by the detector using the reciprocity
theorem.\cite{krcma2025mie, neuman2015} This theorem states that if light can travel from the detector to a point inside the sample, the reverse path is also valid. Therefore, the BLS signal is determined from the interaction between the induced polarization currents in the magnetic film ($\mathbf{P}=\boldsymbol{\chi} \mathbf{E}_\mathrm{dr}$) and an electric field generated at the sample by a virtual source placed at the detector position.

Compared to our previous works,\cite{krcma2025mie, wojewoda2024modeling} we calculate the driving electric field produced by the \textmu{}BLS objective $\tilde{\mathbf{E}}_{\rm dr}$ and the virtual detection field $\tilde{\mathbf{E}}_{\rm v}$ directly in the $k$-space, which leads to significant decrease in the calculation time.\cite{swt} This allows us to obtain the optical transfer function $\boldsymbol{T}$ of the BLS setup via convolution
\begin{equation}
    T_{ij} = \tilde{E}_{{\rm v},i}*\tilde{E}_{{\rm dr},j}.
\end{equation}
The tensor $\boldsymbol{T}$ quantifies the sensitivity of the BLS setup to a range of in-plane wavevectors and is independent of the magnetization dynamics and MO coupling. Finally, the BLS signal $\sigma$ is obtained from the proportionality relation
\begin{equation}
    \sigma(\mathbf{k}_{\rm m},\omega_{\rm m}) 
    \sim 
    \left| \sum_{i,j} \chi_{ij}(\mathbf{k}_{\rm m},\omega_{\rm m}) T_{ij}(\mathbf{k}_{\rm m}) \right|^2,
\end{equation}
where $\mathbf{k}_{\rm m}$ is the in-plane wavevector and $\omega_{\rm m}$ is the angular frequency of the magnon mode.

\begin{figure}[t]
    \includegraphics{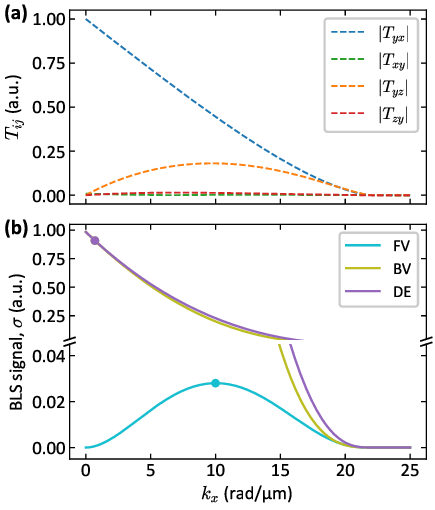}
    \caption{\label{fig:transferfunction} (a)~Absolute values of the transfer function components $T_{yz}, T_{zy}, T_{yx}$, and $T_{xy}$ at $k_y=0$ for positive $k_x$ in the crossed polarizer--analyzer scheme ($\theta=0^\circ$, $\phi=90^\circ$). (b)~BLS signal in the BV, DE, and FV geometry assuming magnetization dynamics with identical amplitudes of the circular spin precession. The spin waves were measured at $k=10$ rad/\textmu{}m for FV and $k=0.7$ rad/\textmu{}m for DE and BV, marked by color dots in the plots.}
\end{figure}

We now illustrate how the spin-wave signal is formed. We assume that the spin waves propagate along the $x$-axis, i.e., perpendicular to the excitation antenna, so that the transfer function can be treated as one-dimensional ($k\equiv k_x$). To simplify the discussion, we consider only the linear Kerr contribution and the crossed polarizer--analyzer configuration, for which the BLS signal is expected to be maximal.

From the structure of the MO tensor in Eq.~(\ref{eq:chi}), there are six non-zero elements. However, the symmetry implies that the $xz$-components contribution cancels, $T_{xz}=T_{zx}=0$. The remaining transfer-function components are shown in Fig.~\ref{fig:transferfunction}(a). The dominating component $T_{yx}$, shown with dashed blue line, is the result of the convolution of the components of the electric field along the polarization direction. This term is responsible for the BLS signal detection in the BV and DE geometries. In the FV geometry, it is not contributing to the BLS signal because $\chi_{yx}=0$. From the remaining components, only $T_{yz}$ plays a significant role. Importantly, this term multiplies with the nonzero susceptibility component $\chi_{yz}=-iQm_x$ in the FV geometry, leading to the possible detection of the BLS signal in this configuration.

The expected BLS signal is shown in Fig.~\ref{fig:transferfunction}(b) in three main spin-wave geometries: FV, BV, and DE. For simplification, we assumed that the magnetization precession is circularly polarized and its amplitude is identical in each geometry. The BLS signal is very similar in BV and DE, having a maximum at $k_x=0$ and decreasing with increasing wavevector. The slight difference comes from the fact that the $T_{yz}$ term of the transfer function contributes to the BLS signal in the DE configuration but not in BV. The signal in the FV geometry is significantly lower than in two other geometries, contributing to its decreased detectability. Interestingly, at $k=0$ the BLS signal is zero, making detection impossible. The maximum of the BLS signal is at the wavevector $k_x=10$~rad/\textmu{}m with a value of only 2.8\% of the maximum BLS signal in DE and BV geometry (at $k=0$).

This simplified (and maximally symmetric) case highlights that the longitudinal component of tightly focused light is the primary contributor to FV spin-wave detectability in \textmu{}BLS. In real experiments, additional pathways may also contribute. For example, allowing for nonzero $k_y$ or including the Cotton--Mouton effect breaks the original symmetry assumptions and can generate the FV signal via $xz$-components of the transfer function. In such cases, the full transfer function must be calculated as 3x3 matrix of 2D reciprocal-space mappings. The complete transfer function of our BLS setup is presented in Section S1 of the Supplementary Material.

\begin{figure*}
    \includegraphics{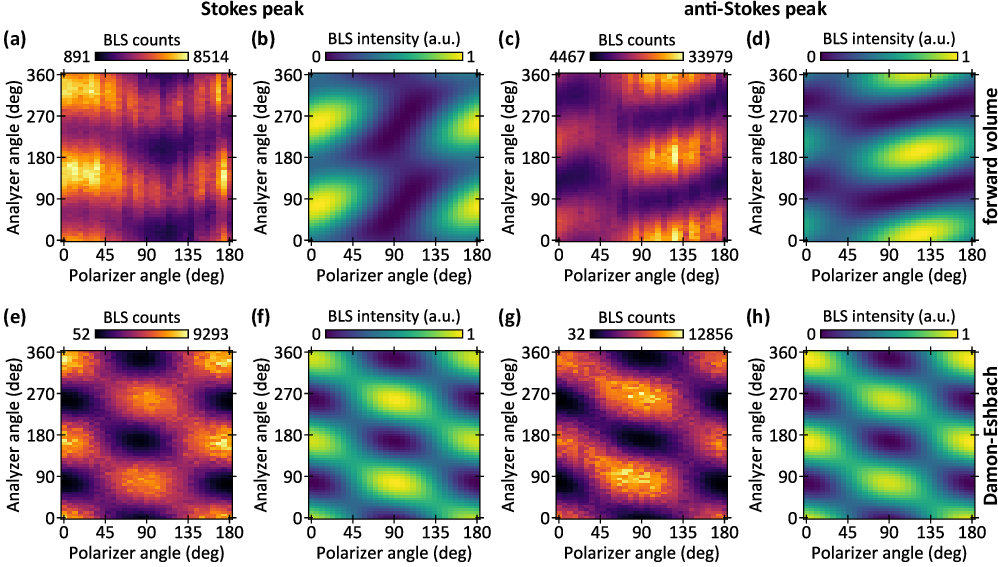}
    \caption{\label{fig:fv-maps} Polarization-dependent BLS amplitude maps of the fundamental magnon mode peak in the experimental data (red colormaps, columns 1 and 3) and fitted using the semi-analytical model (green colormaps, columns 2 and 4). The Stokes (first two columns) and anti-Stokes (last two columns) peaks were analyzed in two fundamental spin-wave geometries: forward volume (a-d) and Damon--Eshbach (e-h).}
\end{figure*}

In the next step, we analyze the full set of the measured spectra. For each polarizer--analyzer angle pair, we integrated the peak intensity and represented each spectrum with a single amplitude value. The resulting amplitudes are shown as colormaps in Fig.~\ref{fig:fv-maps}, separately for the Stokes and anti-Stokes primary peaks in the FV and DE geometries, respectively. The upper and lower half of each plot is almost identical because the $180^\circ$ rotation returns the polarization to its original state. Importantly, all maps clearly indicate a significant contribution of the Cotton--Mouton effect. If the linear Kerr effect dominated, diagonal ridges would be present with maxima at $\phi-\theta=90^\circ$ and $270^\circ$.

The experimentally acquired polarization maps can be reproduced using the semi-analytical model described above with the Cotton--Mouton constants treated as fitting parameters. It is important to note that this polarization analysis allows only to extract the ratio between the Cotton--Mouton constants and the Voigt constant, without the possibility of evaluating the amplitude of any of these constants; therefore, we assume $Q=1$. Prior to fitting the Cotton--Mouton constants, we calibrated the starting positions of the polarizer and analyzer using the elastically scattered signal from the spurious laser mode as a reference (see  Section S2 of the Supplementary Material). In the FV-geometry dataset, this procedure yielded $\phi_0 = 170.86(10)^\circ$, and for the DE it resulted in $\phi_0 = 75.50(14)^\circ$.

After determining the relative polarizer--analyzer angle, we fit the magnon-peak amplitudes. The fit is performed using the least-squares minimization method implemented in \texttt{scipy.curve\_fit} function in Python. Due to the cubic crystal symmetry of BiYIG, we assume $B_{xz}=B_{yz}$. The resulting modeled amplitude maps are shown in Figs.~\ref{fig:fv-maps}(b,d), and the fitted value is $B_{xz}=-0.451(7)+0.195(7)i$.

The model reproduces the anti-Stokes map very well, whereas the agreement for the Stokes map is less accurate. This indicates that some effects cannot yet be fully captured within our semi-analytical description, most likely in the simplified treatment of the MO coupling. Nevertheless, the extracted value of the Cotton--Mouton constant $B_{xz}$ demonstrates that the contribution of the quadratic MO effect in BiYIG is of the same order of magnitude as the linear Kerr effect.

Next, the fitting procedure was repeated for the BV and DE geometries. In these fits, the parameters extracted from the FV dataset were kept fixed, leaving only $B_{xy}$ as a free parameter. The polarization-dependent amplitude maps in DE geometry are presented in Figs.~\ref{fig:fv-maps}(e-h), while the results for BV geometry are presented in Fig.~S4 of the Supplementary Material. In all four cases (Stokes and anti-Stokes peaks in both geometries) in the experiment, the maps show the characteristic checkered pattern. The model reproduces this behavior very well, yielding $B_{xy} = 0.046(3)+1.344(7)i$. 

Together with the FV analysis, these results confirm that the Cotton--Mouton contribution in BiYIG is comparable in magnitude to the linear Kerr effect. More generally, measuring polarization-dependent amplitude maps in multiple field geometries provides a robust route to quantify the relative strength of the Voigt (linear) and Cotton--Mouton (quadratic) MO constants.


In summary, we investigated the feasibility of detecting spin-wave spectra in the FV geometry using \textmu{}BLS. Our semi-analytical model shows that high-NA focusing produces focal field with non-negligible longitudinal electric field component. This component enables MO coupling to the transverse (in-plane) dynamic magnetization in the FV configuration. We experimentally verified these predictions by measuring coherently excited FV spin waves in a thin BiYIG film with \textmu{}BLS. In addition, the measured polarization-dependent spectra reveal clear signatures of quadratic MO coupling (Cotton--Mouton effect). By performing polarization-resolved mapping, we demonstrate that the Cotton--Mouton constants can be quantitatively extracted from \textmu{}BLS data.
More broadly, the ability to map the effective MO tensor in \textmu{}BLS adds dimension to the techniques' capabilities. It offers a route to disentangle overlapping magnon modes by exploiting their distinct polarization fingerprints, separating optical-selection effects from genuine changes in magnon populations, and providing additional constraints for quantitative modeling of mode profiles and hybridization.

\section*{Supplementary Material}
The Supplementary Material contains a detailed description of the transfer function, the polarization-dependent analysis of the laser mode, and polarization-dependent amplitude maps in all three main spin-wave geometries.

\section*{Author contributions}
Krzysztof Szulc: Data curation (lead); Formal analysis (lead); Investigation (lead); Methodology (supporting); Software (supporting); Visualization (lead); Writing -- original draft (equal); Writing -- review \& editing (equal).

Mengying Guo: Data curation (supporting); Formal analysis (supporting); Investigation (lead); Visualization (supporting); Writing -- review \& editing (equal).

Ond\v{r}ej Wojewoda: Conceptualization (equal); Formal analysis (supporting); Investigation (supporting); Methodology (lead); Software (lead); Writing -- review \& editing (equal).

Hongyu Wang: Resources (lead); Writing -- review \& editing (equal).

Dominik Pavelka: Investigation (supporting); Software (supporting); Validation (equal); Writing -- review \& editing (equal).

Jan Kl\'{i}ma: Software (lead); Writing -- review \& editing (equal).

Jakub Kr\v{c}ma: Methodology (lead), Software (lead); Validation (equal); Writing -- review \& editing (equal).

Xiufeng Han: Funding acquisition (equal); Project administration (equal); Resources (equal); Supervision (supporting); Writing -- review \& editing (equal).

Qi Wang: Conceptualization (equal); Funding acquisition (equal); Investigation (supporting); Project administration (equal); Supervision (supporting); Writing -- review \& editing (equal).

Michal Urb\'{a}nek: Conceptualization (equal); Funding acquisition (equal); Investigation (supporting); Methodology (supporting); Project administration (equal); Supervision (lead); Writing -- original draft (equal); Writing -- review \& editing (equal).

\section*{Data availability}
The data that support the findings of this study are openly available in Zenodo at https://doi.org/10.5281/zenodo.18661651.

\begin{acknowledgments}
The authors thank M. Hrto\v{n} for valuable comments regarding reciprocity theorem. This research was supported by project No. CZ.02.01.01/00/22 008/0004594 (TERAFIT). Q.W. acknowledges the financial support of the National Natural Science Foundation of China (Grant No. 12574118). We acknowledge CzechNanoLab Research Infrastructure supported by MEYS CR (LM2023051). This work was financial supported partially by the National Natural Science Foundation of China [NSFC Grant No. T2495210], and the Chinese Academy of Sciences President’s International Fellowship Initiative (PIFI Grant No. 2025PG0006).
\end{acknowledgments}

%

\end{document}



\title{SUPPLEMENTARY MATERIAL\\
       Polarization-resolved measurement of forward volume spin waves by
micro-focused Brillouin light scattering} 



\author{Krzysztof~Szulc}
\email{krzysztof.szulc@ceitec.vutbr.cz}
\affiliation{ 
CEITEC BUT, Brno University of Technology, Brno 612 00, Czechia
}%
\affiliation{ 
Institute of Molecular Physics, Polish Academy of Sciences, Poznań 60-179, Poland
}%
\author{Mengying~Guo}
\affiliation{ 
School of Physics, Hubei Key Laboratory of Gravitation and Quantum Physics, Institute for Quantum Science and Engineering, Huazhong University of Science and Technology, Wuhan, China
}%
\author{Ond\v{r}ej~Wojewoda}
\affiliation{ 
CEITEC BUT, Brno University of Technology, Brno 612 00, Czechia
}%
\affiliation{ 
Department of Materials Science and Engineering, Massachusetts Institute of Technology, Cambridge, MA, USA
}
\author{Hongyu~Wang}
\affiliation{ 
Beijing National Laboratory for Condensed Matter Physics, Institute of Physics, Chinese Academy of Sciences, Beijing, China
}%
\author{Dominik~Pavelka}
\affiliation{ 
CEITEC BUT, Brno University of Technology, Brno 612 00, Czechia
}%
\author{Jan~Kl\'{i}ma}
\affiliation{ 
CEITEC BUT, Brno University of Technology, Brno 612 00, Czechia
}%
\author{Jakub~Kr\v{c}ma}
\affiliation{ 
CEITEC BUT, Brno University of Technology, Brno 612 00, Czechia
}%
\affiliation{ 
Institute of Physical Engineering, Brno University of Technology, Brno 616 69, Czechia
}%
\author{Xiufeng~Han}
\affiliation{ 
Beijing National Laboratory for Condensed Matter Physics, Institute of Physics, Chinese Academy of Sciences, Beijing, China
}%
\author{Qi~Wang}
\affiliation{ 
School of Physics, Hubei Key Laboratory of Gravitation and Quantum Physics, Institute for Quantum Science and Engineering, Huazhong University of Science and Technology, Wuhan, China
}%
\author{Michal~Urb\'{a}nek}
\affiliation{ 
CEITEC BUT, Brno University of Technology, Brno 612 00, Czechia
}%
\affiliation{ 
Institute of Physical Engineering, Brno University of Technology, Brno 616 69, Czechia
}%


\date{\today}

\pacs{}

\maketitle 

\section{Transfer function}

\begin{figure}[!h]
    \includegraphics{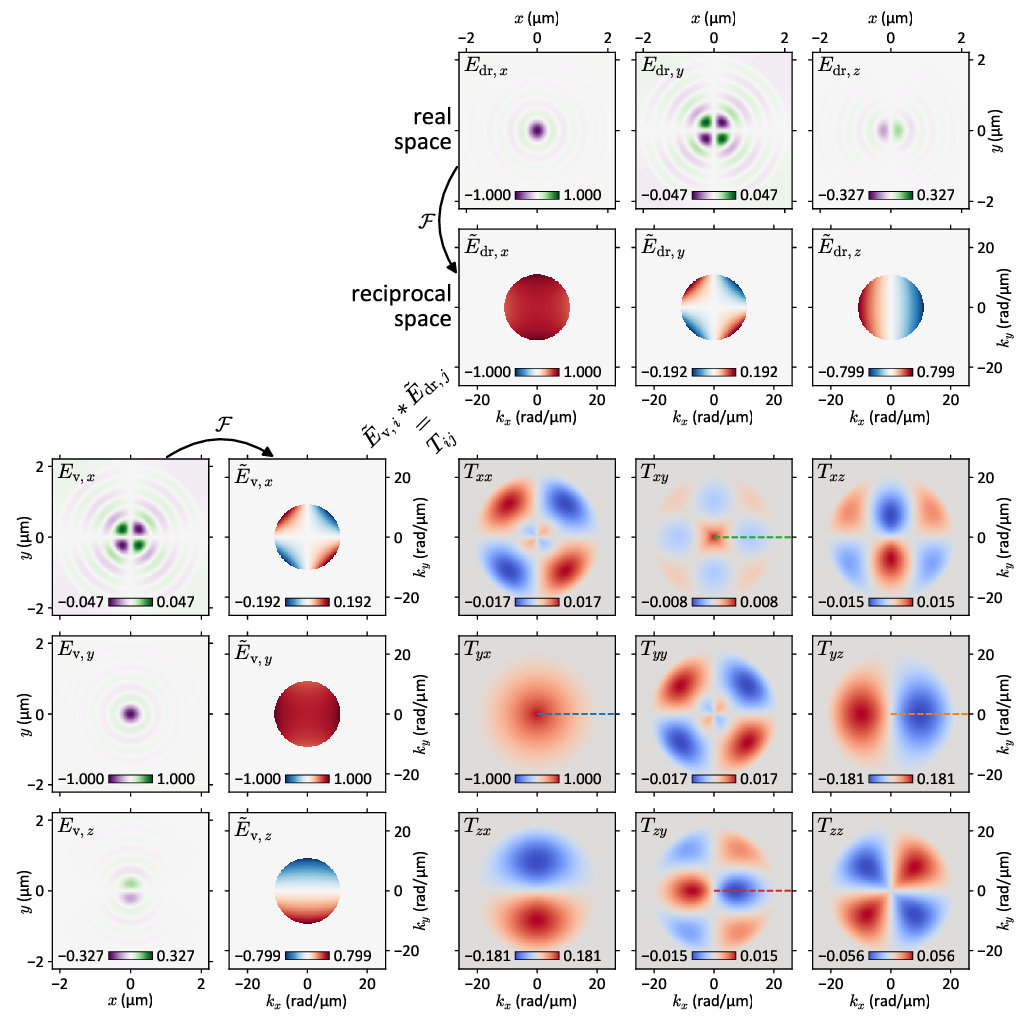}
    \caption{\label{fig:transferfunction} Graphical representation of transfer function calculation. First row from the top and first column from the left show the real part of driving and virtual electric field components in the real space. Second row from the top and second column from left show the driving and virtual electric field components in the reciprocal space. The remaining nine graphs with gray background show the transfer function components. The colorbars on each subplot are normalized to the maximum value of each variable ($\mathbf{E}_{\rm dr},\mathbf{E}_{\rm v},\tilde{\mathbf{E}}_{\rm dr},\tilde{\mathbf{E}}_{\rm v}$ and $\boldsymbol{T}$). Dashed lines in transfer function plots represent the data shown in Fig.~3(a).}
\end{figure}

To better understand the origin of the transfer function, its calculation scheme is shown in Fig.~\ref{fig:transferfunction}. We define our system in the crossed polarizer--analyzer scheme, where incident light has horizontal polarization ($\theta=0^\circ$) and scattered light has vertical polarization ($\phi=90^\circ$). We used the optical setup parameters identical to those used in the manuscript. First, we can look at the electric field in real space. Due to beam focusing, the constriction of the electric field creates small $y$- and $z$-components, which then contribute to the transfer function. Interestingly, apart from Eq.~(3), the transfer function can also be calculated from the electric field in real space using the formula
\begin{equation}
    T_{ij}(\mathbf{k}_m) = \int d\mathbf{r}_\parallel^2 E_{\rm v}^i(\mathbf{r}) E_{\rm dr}^j(\mathbf{r}) e^{i\mathbf{k}_m \cdot \mathbf{r}_\parallel}.
\end{equation}

From the real space, we can transfer the electric field to the reciprocal space by applying the Fourier transform. The characteristic features and symmetries present in the real space are preserved in the reciprocal space. The range of the wavevectors is limited by the numerical aperture of the lens.

Finally, by making the convolution of all components, we obtain the transfer function. It can be clearly understood from this figure how this process leads to the transfer function, which has mixed symmetries from its original components. One can notice, e.g., the similarities between $T_{yz}$ and $T_{zx}$ components. Full information about the transfer function gives us multiple hints about the expected BLS spectra. First, the strongest component $T_{yx}$ comes from the convolution of electric fields along the polarization direction. Because this component multiplies with the dynamic susceptibility component $\chi_{yx}=-iQm_z+B_{xy}m_xm_y$, it is present only when there is an out-of-plane dynamic magnetization component (in the linear regime), and thus when the static magnetization is not perfectly aligned out of the plane. Another important observation is that only the components $T_{yx}$ and $T_{xy}$ have non-zero transfer function at $k=0$. This makes the detection of $k=0$ spin waves in the forward volume geometry impossible.
\newpage

\newpage

\section{Polarization-dependent analysis of the laser mode}

\begin{figure}[b]
    \includegraphics{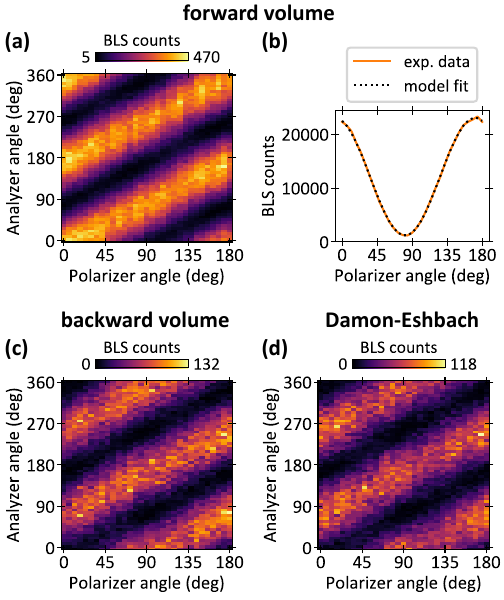}
    \caption{\label{fig:lasermode} Polarization-dependent BLS amplitude maps of the Stokes laser mode peak in the (a) forward volume, (c) backward volume, and (d) Damon--Eshbach geometry. In (b), the fitting is shown for the forward volume geometry.}
\end{figure}

The colormap of the Stokes laser mode peak is shown in Fig.~\ref{fig:lasermode}(a) in the forward volume geometry. The presence of diagonal lines confirms that the peak represents the laser mode. This colormap is transferred to the linear plot shown in Fig.~\ref{fig:lasermode}(b) by summing all points with identical angle between the analyzer and polarizer, precisely, $(\phi-\theta) \mod 180^\circ$. The obtained curve (shown with the solid orange line) is then fitted with the function $y=a\cos{(x-\phi_0)}+b$, where $\phi_0$ represents the real relative angle between the analyzer and the polarizer in their starting position ($\theta=\phi=0^\circ$). In that case, we obtained $\phi_0 = 170.86(10)^\circ$.

Identical fitting procedure was performed for the Stokes laser mode peaks in the backward volume and Damon--Eshbach geometry which polarization-dependent maps are shown in Fig.~\ref{fig:lasermode}(c) and~\ref{fig:lasermode}(d), respectively. For both geometries, we obtained the same angle between polarizer and analyzer $\phi_0 = 75.50(14)^\circ$.

\newpage

\begin{figure}
    \includegraphics{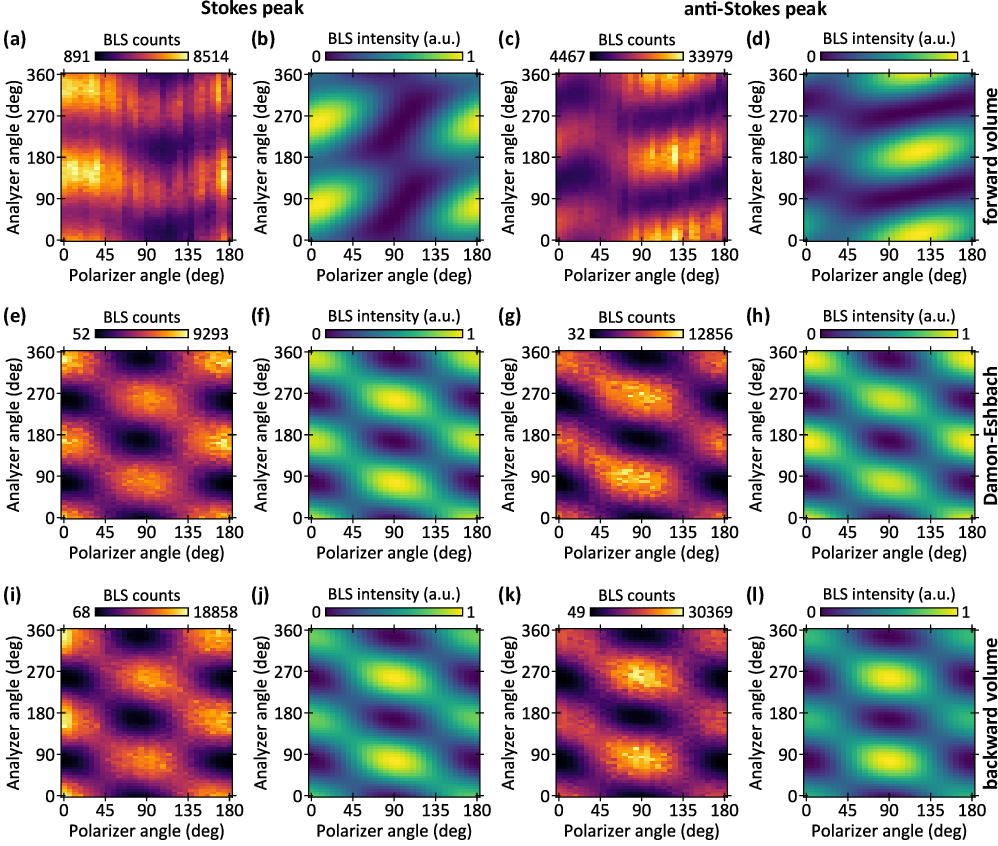}
    \caption{\label{fig:bv-de-maps} (Reproduction of the Fig.~4 with backward volume geometry results). Polarization-dependent BLS amplitude maps of the fundamental magnon mode peak in the experimental data (red colormaps, columns 1 and 3) and fitted using the semi-analytical model (green colormaps, columns 2 and 4). The Stokes (first two columns) and anti-Stokes (last two columns) peaks were analyzed in three fundamental spin-wave geometries: forward volume (a-d), Damon--Eshbach (e-h), and backward volume (i-l).}
\end{figure}


%
%

%
